\documentstyle[11pt,newpasp,twoside,epsf]{article}
\def\te{T_{\rm eff}}
\def\msol{\mbox{M}_\odot}
\def\mjup{\mbox{M}_{jup}}
\def\rjup{\mbox{R}_{jup}}
\def\gcc{\mbox{g}\, \mbox{cm}^{-3}}
\def\simgr{\,\hbox{\hbox{$ > $}\kern -0.8em \lower 1.0ex\hbox{$\sim$}}\,}

\def\wig#1{\mathrel{\hbox{\hbox to 0pt{%
\lower.5ex\hbox{$\sim$}\hss}\raise.4ex\hbox{$#1$}}}}

\begin{document}

\title{The physics of extrasolar gaseous planets : from theory to observable signatures}
\author{G. Chabrier, F. Allard, I. Baraffe}
\affil{C.R.A.L., Ecole Normale Sup\'erieure de Lyon, Lyon, France}
\author{T. Barman}
\affil{Wichita State University}
\author{P.H. Hauschildt}
\affil{Hamburger Sternwarte, Hamburg}

\begin{abstract}
We review our present understanding of the physical properties of substellar objects, brown dwarfs and irradiated or non-irradiated gaseous exoplanets. This includes a description of their internal
properties, mechanical  structure and heat content, their atmospheric properties, thermal profile and emergent spectrum, and their evolution, in particular as irradiated companions of a close parent star. The general theory can be used to make predictions in
term of detectability for the future observational projects.
Special attention is devoted to the evolution of the two presently detected transit planets, HD209458B and OGLE-TR-56B. For this latter, we present a consistent evolution for its recently revised mass and show that we reproduce the observed radius within its error bars. 
We briefly discuss differences between brown dwarfs and gaseous planets, both in terms of mass function and formation process. We outline several arguments to show that the minimum mass for deuterium burning, recently adopted officially as the limit to distinguish the two types of objects,
is unlikely to play any specific role in star formation, so that such a limit is of purely semantic nature and is not supported by a physical justification.
\end{abstract}

\section{Introduction}

Substellar objects (SSO) are defined as objects not massive enough to sustain nuclear fusion in their core, i.e. with mass below the
hydrogen-burning minimum mass ($m_{HBMM}\simeq 0.072\,\msol$, Chabrier \& Baraffe 1997; see Chabrier \& Baraffe 2000, Burrows et al. 2001 for recent reviews).
These once elusive objects have been revealed  to be about as numerous as the Galactic stellar population. Several hundreds of brown dwarfs (BD) have now been discovered in the field or in young clusters and more than a hundred gaseous planets (GP) have been detected outside our solar system.
Many ground-based and space-based searches are planed for the next coming years to seek more of these objects, with as a holy grail the direct observation of an extrasolar planet. These projects require a sound theory to understand the
physics of SSOs and to provide reliable guidance to observation. In this paper, we examine the present status of this theory and confront it to available observations.

\section{Equation of state. Internal structure}

Internal conditions for SSOs, from the HBMM to jovian conditions, range from $T_c\ga 10^6$ K to $10^4$ K and from $\rho_c\sim 10^3\,\gcc$ to $\sim 10\,\gcc$ for the central temperature and density, respectively. Under these conditions, the average potential energy $(Ze)^2/a_i$ of two ions separated by a distance $a_i=3V/4\pi N_i$ can be up to 50 times larger than the average kinetic energy $kT$. This is illustrated by a so-called plasma coupling parameter $\Gamma = (Ze)^2/a_ikT>1$, meaning that the ionic plasma strongly departs from the perfect gas conditions encountered inside massive stars. As a comparison, 
the interior of Sun-like stars is characterized by $\Gamma \sim 0.1$.
Under the aforementioned interior conditions, the Fermi energy of the electrons is comparable    
to the thermal energy of the plasma, $kT_F\approx kT$, so that the electrons form a finite-temperature, partially degenerate plasma. This is illustrated by the so-called degeneracy parameter $\theta = kT/kT_F\sim 0.05$-1. This differs noticeably from for instance the case of white dwarfs, where the electrons are fully degenerate, with $\theta \sim 0.01$. Last but not least, the average binding energy of an electron around a nucleus, $Ze^2/a_i$, is of the order of the electron Fermi energy $\epsilon_F$, so that both H and He atoms experience pressure-ionization, a long standing problem in dense matter physics. The
correct description of the thermodynamic properties of the
strongly correlated, finite-temperature, quantum electron-ion plasma which composes the interior of SSOs thus requires the solution of the so-called N-body problem, one fundamental problem of statistical physics. Until recently, this challenging problem could not be probed by experimental conditions on Earth. Things changed within the past few years with the realization of laser-driven shock wave experiments at Livermore (Collins et al. 1998) and Z-pinch experiments at Sandia (Knudson et al. 2001, 2003) which now reach the megabar range at temperatures characteristic of SSOs, providing stringent constraints on
the internal equation of state (EOS) of these objects. The EOS determines not only the mechanical properties of the object (mass-radius relationship and contraction work) but also its internal heat capacity
and adiabatic gradient. The EOS calculated a decade ago by Saumon \& Chabrier (1992, 1995) has been shown to adequately reproduce the experimental results up to about 0.5 Mbar (Saumon et al. 2000). In the very high density limit, $P\simgr 10$ Mbar, this EOS adequately reproduces Monte Carlo simulations of dense coupled plasmas or first principle expansion schemes (see Saumon \& Chabrier 1992). In the $\sim 0.5$-5 Mbar range, the laser-driven and Z-pinch experiments yied different results, so that more experimental data are needed before the behaviour of hydrogen in this pressure range can be robustly determined.

\section{Atmospheric structure. Emergent spectrum}

Several recent reviews include a description of the spectral evolution of low-mass objects, from  the Sun to jovian planet conditions (Allard et al. 1997, Chabrier \& Baraffe 2000, Burrows et al. 2001).
The atmospheric physical properties of SSOs can be captured from the equilibrium condition of an hydrostatic atmosphere :

\begin{eqnarray}
d\tau=(\rho \bar \kappa_R)\, dz= \bar \kappa ({dP\over g})
\end{eqnarray}

With $\bar \kappa_R \approx 0.01$-1 cm$^2$g$^{-1}$ and $R\sim 0.1$ R$_\odot$, i.e. $g=Gm/R^2\approx 0.1$-10 $g_\odot$, this yields
$P_{ph}\sim g/\bar \kappa \sim$ 0.1-10  bar, $\rho_{ph}\sim 10^{-5}$-$10^{-4}\,\gcc$ at the photosphere,
quite dense conditions for stellar atmosphere standards.
As a function of $\te$, the evolution of the spectral energy distribution can be outlined as follows:

\indent - $\te \la 4000$ K: H$_2$ molecules form. Under the aforementioned density conditions, collisions either between H$_2$ molecules or between H$_2$ and He atoms yield induced dipolar transitions and thus a frequency-dependent, so-called collision-induced absorption (CIA) opacity which scales within first order as the square of the H$_2$ abundance, $\kappa_{CIA}$(H$_2$-H$_2$)$\propto n_{H_2}^2 \kappa_{H_2}$. Besides H$_2$, other molecules form, such as hydrides, FeH, MgH, CaH, metal oxydes, TiO, VO, the dominant sources of absorption in the optical, H$_2$O, CO, which determine the shape of the spectrum in the infrared.

\indent - $\te \la 2500$ K: various refractory compounds (grains) form, such as Mg$_2$SiO$_4$, CaTiO$_3$, MgAl$_2$O$_4$,..., sequestering the metals. These grains yield grey-like absorption in the optical and thus, by energy conservation, lead to a backwarming of the inner atmosphere. This destroys the aforementioned molecules absorbing in the IR, yielding red colors. This lack of metal oxyde (TiO, VO) absorption in the spectrum defines the so-called L-dwarf spectral class. On the other hand, neutral alkali lines (K, Li, Cs, Rb, Cs) in the optical and near-IR become increasingly important.

\indent - $\te \la 1800$ K. In this regime, the dominant equilibrium form of carbon becomes CH$_4$, at the
expense of CO. The absorption of methane in the H and K bands yields a redistribution of the flux at shorter
wavelengths, yielding bluer colors. This presence of methane in the spectrum defines the T-dwarf spectral type, which encompasses both cool BDs and GPs. The benchmark of this type of objects, with
jovian-like atmospheres, is the first unambiguously identified BD, Gl229B (Oppenheimer et al. 1995), of which spectral energy distribution was successfully
predicted by the theory (Allard et al. 1996, Marley et al. 1996).

\indent - $\te \la 500$ K. We enter the conditions characteristic of the jovian planets of our solar system, reaching the condensation limits of H$_2$0 and then NH$_3$.

Figure 1 displays the spectrum of a low-mass M-dwarf with $\te=2500$ K, a L-type BD with $\te=1800$ K
and a T-dwarf BD with $\te=1000$ K, typical of non-irradiated GP conditions, from Allard et al. (2001) atmosphere models. This series of spectra illustrates the aforedescribed temperature evolution of the dominant sources of absorption. The role of grains in the L-dwarf spectrum is obvious from the grey-like behaviour at short wavelengths, due dominantly to Mie scattering, and
the reduced bands of H$_2$0 and CO in the IR, compared with the hotter spectrum. The constant replenishment of these grains in the atmosphere is
due to convective motions which reach the regions near the photosphere (Ludwig, Allard \& Hauschildt 2002). For cooler temperatures and luminosities, characteristic of T-dwarfs and planets, most grains form below the photosphere, convection
retreats to deeper layers and the largest grains sink under the action of gravity
to deeper, hotter regions where they melt and sublimate. 

\begin{figure}
\epsfxsize=80mm
\epsfysize=70mm
\plotone{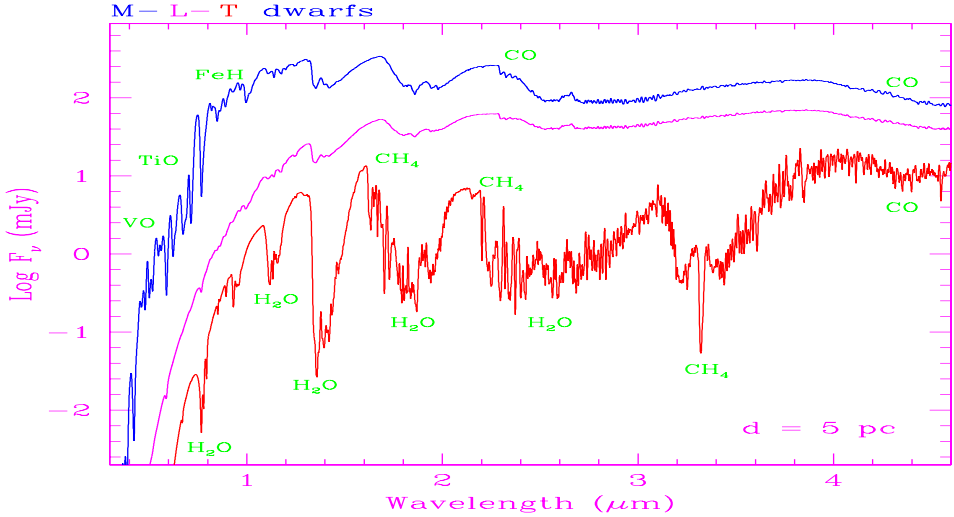}
\caption{Spectral energy distribution of a typical M-dwarf ($\te=2500$ K), L-dwarf ($\te=1800$ K) and T-dwarf ($\te=1000$ K), from top to bottom, from Allard et al. (2001). Surface gravities and radii are from Chabrier et al. (2000a).}
\label{fig1.ps}
\end{figure}

\section{Evolution}

Given the strong frequency dependence of the emergent spectrum of SSOs, and possible convective instability in the atmosphere, correct evolutionary calculations require a consistent (non-grey) treatment between the internal structure profile
and the atmospheric profile, as described in Chabrier \& Baraffe (1997). A grid of atmosphere models with various $(\log g, \te)$ conditions is computed, and matched to
the interior profile at deep enough optical depth ($\tau \approx 100$) to reach the internal adiabat.
For a given mass and age ($m,t$), there is a unique solution matching the two profiles, yielding
consistency between the observable signature, spectrum and magnitude, and the fundamental properties of the object, $m,\,R,\,\te,$ and $t$. The evolution obeys the usual first and second principles of thermodynamics:

\begin{eqnarray}
L=4\pi R^2\sigma \te^4=-\int ({du \over dt} - {P\over \rho^2}{d\rho \over dt})\, dm\,
=\, \int -T {d\tilde S\over dt}dm
\label{evol}
\end{eqnarray} 

\noindent where $\tilde S$ denotes the specific entropy of the BD or the planet and $R$ its radius.

The evolution of the mass-radius relationship in the stellar and substellar regime is shown in Figure 3 of Chabrier \& Baraffe (2000).
The relatively constant behaviour of the radius as a function of mass in the substellar regime stems
from the compensating contributions of the quantum electron gas, which yields an increasing radius
with decreasing mass, a well-known property of degenerate matter, and of the classical, strongly correlated ionic plasma. This
yields a maximum radius near 4 $\mjup$ (Hubbard 1994, Chabrier \& Baraffe 2000). This $m$-$R$ relation has been confirmed recently in the very-low-mass star range by the
remarkable agreement with the radii measured by interferometry (Lane, Boden \& Kulkarni 2001, S\'egransan et al. 2003). On the other hand, recent observations of the radii of the eclipsing binary YY-Gem, with
masses $m_1$=$m_2$=$0.6\,\msol$ (Torres \& Ribas 2002) show a $\sim 10\%$ discrepancy between the theoretical and observed radii. These
stars, however, show intense signatures of activity, a common feature for very short
period systems such as eclipsing binaries. The surface, indeed, is observed to be covered by magnetic spots, reducing the total irradiating area. Calculations in the case of cataclysmic variables (Renvoiz\'e et al. 2002) have shown that a reduced radiating surface leads to a reduced contraction at early stages of evolution, and thus a larger radius. Similar effects may apply to active, short-period systems such as YY Gem.

Evolutionary sequences $L(t)$, $\te(t)$ and color-magnitude diagrams for SSOs down to $0.5\times 10^{-3}\msol=0.5\, \mjup$ are presented  in Baraffe et al. (2003). In a CMD,
the $\te$-sequence of absorbers described in \S3 leads to redward and blueward zig-zag behaviours along an isochrone.

\section{Effect of irradiation}

The majority of the presently observed exoplanets are in very close orbit to their
parent star. In that case, the incoming stellar radiation modifies substantially not only the emergent spectrum but also the evolution of the
planet, as shown initially by Guillot et al. (1996). These modifications are examined below.

\subsection{Atmosphere}

Under strong irradiation conditions, the solution of the transfer equation, which determines the atmospheric
thermal structure of the planet, must include the incoming incident stellar flux in the source function.
Such calculations were first conducted by Seager \& Sasselov (1998), using a limited set of opacities, and
have been improved by Barman et al. (2001)
and Sudarsky et al. (2003).
Figure 5 of Baraffe et al. (2003) compares the thermal structure of two planets with intrinsic temperatures $\te=1000$ K and $\te=100$ K, respectively, orbiting
at 0.05 AU of a G2 star, when irradiation is or is not taken into account, as calculated in Barman et al. (2001). The irradiated thermal profile is shown to be strongly modified with respect to the non-irradiated one. 
Heating of the upper layers of the atmosphere leads to a nearly-isothermal external profile, yielding eventually a temperature inversion in the outermost layers\footnote{Note that the outermost temperature inversion
observed in Jupiter, which affects the UV part of the spectrum, is due to chromospheric effects, not included in the present calculations. These upper atmosphere
effects, however, do not bear consequences on the evolution.}.
As a consequence of this external heating, the radiative zone of the atmosphere extends to deeper layers, so that the radiative-convective boundary retreats towards the interior.
This demonstrates the necessity to do {\it consistent} calculations between the irradiated atmosphere profile and the inner profile. Matching arbitrarily the inner profile to an atmospheric profile defined by
the equilibrium temperature of the planet, as done in previous evolutionary calculations, yields severely flawed results.

\begin{figure}
\epsfxsize=80mm
\epsfysize=70mm
\plotone{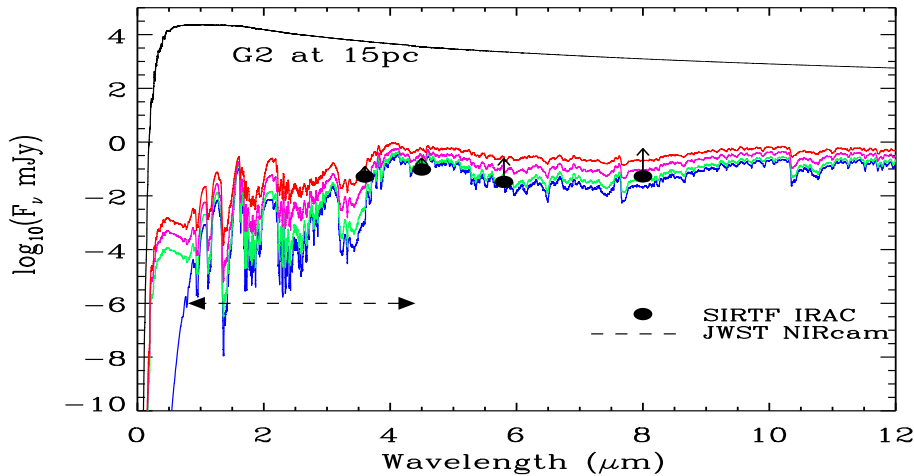}
\caption{Spectrum of a planet with intrinsic temperature $\te = 500$ K, orbiting a G2 star at 0.3, 0.5 and 1.0 AU, respectively, from top to bottom, from Barman et al. (2001). The bottom curve displays the non-irradiated spectrum of the planet. These calculations correspond to a cloudless COND atmosphere model}
\label{fig2.ps}
\end{figure}

Figure 2 displays the emergent spectrum of a $\te = 500$ K, $\log g=3.5$ planet orbiting at 0.3, 0.5 and 1.0 AU from its G2 parent star. The infrared part of the spectrum displays the intrinsic, thermal spectrum of the planet and exhibits the absorption features mentioned in \S3. These features,
however, are less pronounced than in the non-irradiated case, as seen from the comparison with the
bottom line in the figure, a consequence of the aforementioned flat atmosphere temperature profile. This figure displays the calculations done with the so-called COND models, where grains are supposed to have sunk below the photosphere. This case maximizes the effect of irradiation,
for the impending radiation penetrates deeper into the atmosphere than in the case of the so-called
DUSTY models, where cloud layers form above the photosphere (see Barman et al. 2001). The short wavelength part of the spectrum corresponds to the reflected light from the parent star and clearly
mirrors the absorption features of this latter. At short wavelengths, the albedo is dominated by  H or H$_2$ and He Rayleigh scattering. Because
of the "grey" characteristics of dust opacity, the reflected spectrum in the case of the DUSTY models
displays fewer spectral features (Barman et al. 2001). Detectability limits of the SIRTF and JWST missions are indicated in the figure for illustration. Note also the contrast between the planet and the star flux, which goes from $\sim 10^{-3}$ in the IR to larger values shortward of $\sim$4 $\mu$m.

Note that the present calculations assume that all the incident flux is deposited at the substellar
point, namely the planet point closest to the star, and that the energy is concentrated on the day-side
of the planet. Such assumptions maximise the effect of irradiation.

\subsection{Evolution}

First consistent evolution calculations of irradiated planets have been conducted by Baraffe et al. (2003) and more recently by Burrows et al. (2003).
The calculations proceed as described earlier for the non-irradiated case. A grid of irradiated atmospheric structures
is computed for various $(\log g, \te)$ conditions, for a given orbital distance $a$, and interpolation between these structures yield the unique boundary condition with the internal structure.
As mentioned above, irradiation pushes the radiative-boundary limit further deep in the interior, so
that, under strong irradiation conditions, the boundaty condition between the atmospheric and internal
structure may lie in a radiative layer. For this reason, we have computed the Rosseland mean of the atmospheric opacities and used this value to calculate the radiative
gradient for the interior structure. The incoming radiative flux $F_{inc}={1/2 }({R_\star / a})F_\star$ modifies the energy budget of the planet and the emergent total flux $F_{out}$, which now reads (Baraffe et al. 2003):

\begin{eqnarray}
F_{out}=F_{inc}+\sigma \te^4 
=A\,F_{inc} + (1-A)\,F_{inc} + \sigma\,\te^4 
\label{evolir}
\end{eqnarray} 

\noindent where $A$ denotes the Bond Albedo and the first term on the r.h.s. of (3) describes the reflected part of the spectrum. We stress that the albedo $A$ is not a free parameter but is calculated consistently with the radiative transfer equation. This yields for the total luminosity:

\begin{eqnarray}
L_{tot}=L_{reflected} \,+\,4\pi R^2_p\sigma (T_{eq}^4+\te^4)
\,=L_{reflected} + 4\pi R^2_p\sigma T_{eq}^4 - \int T {d\tilde S\over dt}dm
\label{evolirrad}
\end{eqnarray} 

Note that only the last term on the right hand side of eq.(4) concerns the evolution of the
intrinsic luminosity of the planet $L_{int}$. The second term defines its equilibrium temperature, i.e. the temperature the planet would reach in the absence of any internal source of energy, including from its contraction. The first term, which defines the Bond albedo,
illustrates the luminosity fraction of the parent star reflected by the planet atmosphere.
The dominant contribution to $L_{int}$, both in the irradiated and non-irradiated cases,
arises from the contraction work contribution, $ \int_M  {P\over \rho^2}{d\rho \over dt}\, dm=-\int_M  {Gm\over r^2}{dr \over dt}\, dm$.

\begin{figure}
\epsfxsize=60mm
\epsfysize=60mm
\plotone{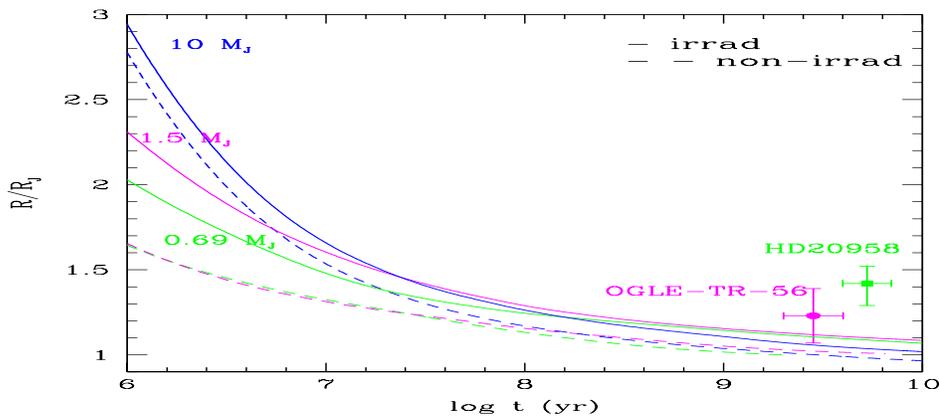}
\caption{Evolution of non-irradiated (dash) and irradiated (solid) gaseous planets orbiting a G2 star
with $T_{eff_\star}=6000$ K. The $0.69\,\mjup$
and $10\,\mjup$ planets are located at 0.05 AU, while the $1.5\,\mjup$ is at 0.023 AU}
\end{figure}

Figure 3 illustrates the evolution of the radius for different planets orbiting a G2 star, including the two detected transit planets HD209458B and OGLE-TR-56B.
At a given orbital distance, namely $a=$0.05 AU, the effect of irradiation is smaller for the more massive, and thus hotter 10 $\mjup$ planet, compared with the 0.69 $\mjup$ one (HD209458B).
This was expected from the smaller modification of the inner atmosphere profile due to a smaller stellar/planet flux ratio (see Fig. 5 of Baraffe et al. 2003). Irradiation, however, slows down
substantially the contraction of the less massive planet, in particular at the early stages of evolution,
when degeneracy effects in the interior are smaller. The figure displays also the evolution of OGLE-TR-56B, for its recently {\it revised} mass, $m\simeq 1.45\pm0.23\,\mjup$, for a radius $1.23\pm 0.15\,\rjup$ (Torres et al. 2003) and an orbital distance $\sim 0.023$ AU. The observed radii of the two transits are shown in the figure. The theoretical radius lies clearly below the observed one for HD209458B, but fits within the error bar for OGLE-TR-56B for the age of the system.
We derive $m\simeq 1.05\,\msol$, $t\sim 4$ Gyr to reproduce the observed extinction-corrected magnitude of the parent star OGLE-TR-56, in agreement with Sasselov (2003).
These results show that there is presently no inconsistency between theory and observation for this transit (assuming OGLE-TR-56B is indeed a real transit).

 As discussed at length in Baraffe et al. (2003), however, no consistent evolution calculation can reproduce the observation of HD209458B within the error bars. Note that our calculations take into account the extension of the atmosphere, which we find to be of the order of $\sim 0.05\,\rjup$ (at optical wavelengths) for the age of the object. Given the remaining uncertainties
in the opacities of these planets, which should include processes such as non-equilibrium chemistry,
dynamics of grains, etc..., we cannot exclude larger extensions. However, as mentioned above, the present calculations maximize the effect of irradiation, and it is very unlikely that different opacities modify the internal adiabat strongly enough
to get a further 30\%
increase of the planet radius. Such a modification requires a dramatic change of the energy
deposited at the top of the internal adiabat, which determines the radius. But either molecular
absorption or, at hotter temperatures, bound-free transitions and free-free scattering of the electrons in the interior prevent photons to penetrate deep enough.

As discussed in \S4.3 of Baraffe et al. (2003) and in Guillot \& Showman (2002), one needs an extra source of energy to be deposited
early enough in the planet, in the first $\sim10^7$ years, and to remain constant after this stage, in order
to reproduce the radius of HD209458B. Tidal dissipation or synchronization are unlikely, for these timescales
are at most of the order of $10^8$ yr.  Note in passing that Bodenheimer et al. (2003) point out the possibility of a second
planet orbiting HD209458B. This constant tidal heating could provide the required mysterious extra source of energy.

\section{Brown dwarfs versus planets}

The deuterium-burning minimum mass, $m_{DBMM}\simeq 0.12\,\msol$ (Saumon et al. 1996, Chabrier et al. 2000b) has been designated recently by the IAU as the official limit between BDs and planets. Although any definition can be accepted on a purely semantic ground, it is important to recall why the DBMM is supposed to play a particular role in star formation, distinguishing eventually
two different formation mechanisms. This issue, as various star formation scenarios, is discussed at length in the review by Chabrier (2003) and will be only shortly summarized below.

The particular role of the DBMM comes from the canonical star formation theory of Shu, Adams \& Lizano (1987). In this theory, low-mass stars form from the collapse of initially {\it hydrostatic}, unstable
dense cloud cores which have reached a $\rho(r) \propto r^{-2}$ density distribution of a singular isothermal spheroid. In this scenario, D-burning at pre-main-sequence ages
is the key ingredient to trigger star formation. The onset of D-burning induces
convective instability in the interior. Combined with the rapid rotation resulting from the accretion of
angular momentum with mass, this convection is believed to generate a strong magnetic field through
dynamo process. The field will drive a magnetocentrifugal wind that ultimately
sweeps away the surrounding accreting material, and determines the mass of the nascent star.
Within this picture, objects below the D-burning limit can not reverse the infall and thus can not form gravitationally bound objects. In this picture, the ambipolar diffusion timescale, $\tau_{ad}\ga 10$ Myr for a cloud of average density $\sim 10^2$ cm$^{-3}$ and $B\approx$ a few $ \mu$G, sets up the characteristic time scale for star formation. Such a scenario can now be reasonably
excluded for several reasons: (1) SSOs are fully convective, with or without D-burning (Chabrier \& Baraffe 2000, Chabrier et al. 2000a), (2) star formation is a rapid process, more rapid than
the thermal crossing timescale or the aforementioned ambipolar diffusion timescale, and comparable to
the dynamical timescale $\tau_{dyn}=(3\pi/ 32G\rho)^{1/2} \approx$1-5$\times 10^5$ yr for typical star-forming molecular clouds, (3) in the ambipolar diffusion scenario, the cloud must survive long enough in near equilibrium between magnetic and gravitational pressure.
This is not consistent with observations of rapid star formation and cloud dissipation, and with the observed turbulent nature of clouds. Indeed, equipartition between kinetic, gravitational and magnetic energy fails to reproduce the observed
properties of molecular clouds, dominated by super-Alfv\'enic and supersonic motions, where kinetic energy dominates magnetic energy, with a decay timescale approximately equal to a dynamical timescale. Moreover, ambipolar diffusion models require large static magnetic field strengths ($\sim$30-100 $\mu$G) exceeding
Zeeman estimates for low-mass dense cores (see e.g. Padoan \& Nordlund 1999 and references therein). This indicates that, although magnetic field probably plays some role in star formation, it is unlikely to be a dominant process.
On these grounds, D-burning seems to be excluded as
a peculiar process in star formation. This is supported observationally by the numerous free-floating objects which have been detected below the DBMM (Zapatero Osorio et al. 2000), although one must remain cautious about the age determination of the cluster and the related mass determination from the theoretical models (see Baraffe et al. 2002). In conclusion, the choice of the DBMM as the mass limit
to distinguish BDs from planets
does not appear to be supported by physical considerations and must be considered as a purely semantic definition.

A more physically-grounded minimum mass for star formation is the opacity-limit mass for fragmentation. This defines the theoretical minimum mass for cloud collapse between the initial, low-density phase, where compressional heating is balanced by molecular line emission, with $P\propto \rho$, and the subsequent, high-density ($\rho_c\simgr 10^{13}\gcc$ for a cloud temperature $T\sim 10$ K) adiabatic phase where the gas becomes optically thick to infrared radiation, with $P\propto \rho^{5/3}$ (Larson 1969). This sets a minimum mass for fragmentation of the order of Jupiter mass (Boss 2001 and references therein).
Revealing observationally the existence of this limit would be a tremendous breakthrough. 

The triggering
process for star formation seems to be the small scale dissipation of large scale supersonic turbulence. This process leads to the formation of dense cores, some of which becoming dense enough
for pressure forces to balance gravity (Padoan \& Nordlund 2002). These cores eventually fragment further to form multiple systems (see e.g. Delgado-Donate et al. 2003, Goodwin et al. 2003). In terms of mass function, the mass distribution of the cores thus corresponds more or less to the IMF of {\it systems}, whereas further fragmentation leads to the IMF of individual objects. As shown by Chabrier (2003, figures 3-5), the present observations of BDs either in the field or in young clusters are consistent (within possibly a factor of 2 to 3, depending on the real BD binary fraction) with the extension into the substellar regime of the same stellar single object or system IMF.
In the opposite, the mass distribution of exoplanets differs drastically from such an IMF. 
A statistical analysis of their
mass distribution, corrected for the uncertainty due to the inclination $sin \, i$ of the orbital plane on the sky, has been
established by Zucker \& Mazeh (2001) and Jorissen, Mayor \& Udry (2001). The resulting planetary mass  distribution  $dN/dm_p$
peaks around $\sim$ 1-2 $M_{Jup}$ - significantly below the DBMM - due to present detection limits of radial velocity surveys to detect smaller objects, and decreases rapidly with only a few objects above $\sim 10\,M_{Jup}$. This distribution corresponds to
a relatively flat MF, $\xi(\log m)\approx$cst, below $\sim 10\,M_{Jup}$.
This different mass distribution
clearly points to a different population of SSOs, namely {\it planetary} companions of stars, and
thus to a formation mechanism different from the one which yields the stellar+BD IMF. 
The only identified difference between GPs like Jupiter or Saturn and BDs
is the presence of a rock+ice core, of several Earth masses, at the very center of these planets. This oversolar average abundance of heavy elements for our jovian planets suggests the presence of a disk in
their formation environment and thus differs from the star+BD collapse scenario. Planets are thus unlikely to form as isolated, free floating objects, although some could have been ejected. On the other hand, the significant fraction of BDs in young clusters
observed to harbor near and mid-IR excess consistent with dusty disks points to a formation mechanism for BDs similar to the standard cloud collapse star formation process (Jayawardhana et al. 2003). The only {\it indirect} clue about the presence of a rocky core in our jovian planets, however, stems from the very accurate determinations of their
gravitational moments from the Voyager, Pionneer and Galileo missions. Such data are obviously unavailable for the exoplanets (although transit observations may provide some information about the presence of a core) and for the moment the only possible way to distinguish BDs from planets is a statistical argument, with two radically different mass distributions. Therefore, the observed so-called free floating "planets" or "sub brown dwarfs" represent most likely the low-mass tail distribution of usual brown dwarfs. In any case, the search for such very-low mass objects is of tremenduous interest for it will allow us to identify the minimum mass for star formation.

\section{Conclusion and perspective}

This review outlines our present understanding of the physics of substellar objects, brown dwarfs and
irradiated or non-irradiated gaseous planets. Various successes of the theory, which had correctly {\it predicted} different mechanical and thermal properties such as the EOS up to the megabar range, the spectral energy distribution of Gl-229B like objects, dominated by methane absorption, the photometric evolution of dusty and dustless
atmosphere SSOs, and the mass-radius and mass-magnitude relationships in the stellar range, give us confidence
in the reliability of its basic foundations, although there is certainly room for a lot of improvements. As argued in this review,
a consistent treatment between the atmospheric profile and the interior profile is mandatory to yield
reliable results, both in the irradiated and non-irradiated case. This theory allows us to derive the fundamental properties of an object (mass, temperature, age, radius) from its observable spectroscopic or photometric signatures. Conversely, the theory provides reliable basis to guide observational projects. Our understanding of the
properties of strongly irradiated planets, the so-called hot jupiters, is still questioned. Whereas the theory adequately reproduces the radius of OGLE-TR-56B, no calculation can reproduce the one of HD209458B. Whether
this latter is a common representative of hot-jupiter objects or whether its large radius is due to tidal interaction with an undetected companion needs to be assessed.
Ongoing searches for transit planets will help to answer this question and confirm or not our present description of strongly irradiated planets.

\end{document}